\documentclass[10pt,conference]{IEEEtran}
\IEEEoverridecommandlockouts
\usepackage{adjustbox}
\usepackage{algorithm}
\usepackage{algorithmic}
\usepackage{amsfonts}
\usepackage{amsmath}
\usepackage{amssymb}
\usepackage{array}
\usepackage{balance}
\usepackage{colortbl}
\usepackage{diagbox}
\usepackage{float}
\usepackage{graphicx}
\usepackage{hyperref}
\usepackage[utf8]{inputenc}
\usepackage{lipsum}
\usepackage{makecell}
\usepackage{multirow}
\usepackage{pgf-pie}
\usepackage{pgfplots}
\usepackage{pgfplotstable}
\usepackage{slashbox}
\usepackage{tabularx}
\usepackage{textcomp}
\usepackage{tcolorbox}
\usepackage{tikz}
\usepackage{xcolor}
\usepackage{xurl}
\pgfplotsset{compat=1.18}
\usepgfplotslibrary{polar}
\usetikzlibrary{patterns, shapes.geometric, arrows, positioning, calc}

\makeatletter
\newcommand{\linebreakand}{%
  \end{@IEEEauthorhalign}
  \hfill\mbox{}\par
  \mbox{}\hfill\begin{@IEEEauthorhalign}
}
\makeatother

\begin{document}

\title{From HNSW to Information-Theoretic Binarization: Rethinking the Architecture of Scalable Vector Search}


\author{
    \IEEEauthorblockN{Seyed Moein Abtahi}
    \IEEEauthorblockA{%
    \textit{Faculty of Engineering and Applied Science}\\
    \textit{Ontario Tech University}\\
    seyedmoein.abtahi@ontariotechu.ca}
    \and
    \IEEEauthorblockN{Majid Fekri}
    \IEEEauthorblockA{%
    \textit{Moorcheh AI}\\
    \textit{EdgeAI Innovations}\\
    majid.fekri@edgeaiinnovations.com}
    \and
    \IEEEauthorblockN{Tara Khani}
    \IEEEauthorblockA{%
    \textit{Moorcheh AI}\\
    \textit{EdgeAI Innovations}\\
    tara.khani@edgeaiinnovations.com}
    \linebreakand 
    \IEEEauthorblockN{Akramul Azim}
    \IEEEauthorblockA{
    \textit{Faculty of Engineering and Applied Science}\\
    \textit{Ontario Tech University}\\
    akramul.azim@ontariotechu.ca}
}

\maketitle

\begin{abstract}
Modern semantic search and retrieval-augmented generation (RAG) systems rely predominantly on in-memory approximate nearest neighbor (ANN) indexes over high-precision floating-point vectors, resulting in escalating operational cost and inherent trade-offs between latency, throughput, and retrieval accuracy. This paper analyzes the architectural limitations of the dominant “HNSW + float32 + cosine similarity” stack and evaluates existing cost-reduction strategies, including storage disaggregation and lossy vector quantization, which inevitably sacrifice either performance or accuracy. We introduce and empirically evaluate an alternative information-theoretic architecture based on maximally informative binarization (MIB), efficient bitwise distance metrics, and an information-theoretic scoring (ITS) mechanism. Unlike conventional ANN systems, this approach enables exhaustive search over compact binary representations, allowing deterministic retrieval and eliminating accuracy degradation under high query concurrency. Using the MAIR benchmark across 14 datasets and 10,038 queries, we compare this architecture against Elasticsearch, Pinecone, PGVector, and Qdrant. Results demonstrate retrieval quality comparable to full-precision systems, while achieving substantially lower latency and maintaining constant throughput at high request rates. We show that this architectural shift enables a truly serverless, cost-per-query deployment model, challenging the necessity of large in-memory ANN indexes for high-quality semantic search.
\end{abstract}


\section{Introduction }
\label{sec:introduction}

\subsection{The Vector Search Cost-Performance Problem}
Modern RAG systems and autonomous agents need semantic search with millisecond latency. Vector databases like Pinecone, Qdrant, Weaviate, Milvus, and Redis have all chosen the same approach: keep everything in RAM. But this creates a real problem. More vectors means more RAM, and RAM is expensive. The more your data grows, the more your costs grow. In this paper, we look at how today's systems work, why they're expensive, what companies have tried to fix it, and a new approach that might change everything.

\subsection{Architectural Foundation and Cost Implications}
\label{subsec:architectural_paradigm}

RAG systems need to search vectors quickly. They need results in milliseconds, not seconds, because search is blocking: the system waits for results before moving forward \cite{digitalocean_choose_vector_db_rag}. When search is slow, everything else breaks down. The answers get worse and the whole system fails more often \cite{zenml_tested_vector_dbs_rag}.

To get millisecond results, everything has to be in RAM. If vectors live on disk, you have to read from disk, and that's too slow for real-time search \cite{sonia_choosing_right_vector_db_arch, zilliz_vector_vs_inmemory}. So the choice is forced: keep everything in memory. But this ties your costs directly to how much RAM you need. The bigger your dataset, the more you spend.

The math is simple and brutal. One billion vectors, each 768 dimensions, stored as float32 takes about 3 terabytes of RAM \cite{jo_hybrid_hnsw_if}. That's a lot of expensive hardware. When you scale to bigger datasets, your main cost is buying more RAM. There's no way around it \cite{zilliz_vector_vs_inmemory}.

\subsection{The Three-Component Algorithmic Stack}
\label{subsec:algorithmic_stack}

Every vector database does the same three things. First, it uses HNSW indexing to find nearby vectors fast. Second, it stores vectors as float32 numbers to keep precision high. Third, it measures distance between vectors using cosine similarity. Each of these choices affects both speed and cost.

\subsubsection{Index Structure: HNSW}
\label{subsubsec:hnsw}

HNSW represents the state-of-the-art technique for approximate nearest neighbor (ANN) search \cite{arxiv_vector_db_survey}. This graph-based indexing structure employs a hierarchical topology enabling efficient location of semantically similar vectors through greedy traversal \cite{mdpi_p_hnsw, arxiv_vector_db_survey}. The algorithmic efficiency derives from superior recall-latency trade-offs; however, performance is contingent upon complete graph structure residency within RAM \cite{arxiv_d_hnsw}, establishing a non-negotiable coupling between index cardinality and required memory capacity.

\subsubsection{Numerical Representation: Float32 Vectors}
\label{subsubsec:float32}

Vector embeddings, high-dimensional numerical representations of unstructured data \cite{jo_hybrid_hnsw_if}, are standardly encoded using single-precision floating-point (float32) representation, requiring 32 bits of storage per dimension \cite{mdpi_p_hnsw}. This high-precision format facilitates accurate distance calculations but substantially increases memory requirements.

\subsubsection{Distance Metric: Cosine Similarity}
\label{subsubsec:cosine_similarity}

Semantic search requires computing inter-vector similarity via cosine similarity, which quantifies the angular relationship between vector pairs:

\begin{equation}
\text{sim}(\mathbf{A}, \mathbf{B}) = \frac{\mathbf{A} \cdot \mathbf{B}}{\|\mathbf{A}\|_2 \cdot \|\mathbf{B}\|_2}
\label{eq:cosine_similarity}
\end{equation}

where $\mathbf{A}, \mathbf{B} \in \mathbb{R}^d$ represent $d$-dimensional embeddings \cite{ibm_cosine_similarity}. Computational complexity scales linearly with embedding dimensionality, requiring substantial floating-point arithmetic operations (multiplication and accumulation), resulting in persistent high CPU utilization \cite{tigerdata_cosine_python}.

\subsection{Cost Structure Analysis}
\label{subsec:cost_analysis}

Integration of HNSW, float32 representation, and cosine similarity generates a two-pronged cost structure. HNSW's in-memory graph traversal requirement necessitates continuous, high-capacity RAM provisioning for both index and full-precision vectors \cite{jo_hybrid_hnsw_if, cloudnative_vector_db_landscape}. Simultaneously, cosine similarity computation on high-dimensional float32 vectors imposes continuous, elevated computational load, generating sustained CPU resource consumption \cite{ibm_cosine_similarity, tigerdata_cosine_python}.

The problem is that you can't change one part without breaking everything else. If you move vectors to cheaper disk storage, you break HNSW's requirement to have everything in RAM, and your search gets too slow \cite{aws_s3_vectors_tiered_storage}. So, the only way to reduce costs is to replace all three parts at once with a fundamentally different execution model, including reconsideration of approximate indexing itself.

\section{Existing Cost Mitigation Strategies: Compromise Architectures}
\label{sec:compromise_architectures}

Faced with prohibitive total cost of ownership (TCO) in pure in-memory deployments, the vector database market has introduced architectures reducing costs through systematic sacrifices in either latency or accuracy.

\subsection{Storage Disaggregation: The Latency-Cost Trade-off}
\label{subsec:storage_disaggregation}

Storage disaggregation implements a hybrid topology, decoupling compressed index structures from full-precision vector data \cite{sonia_choosing_right_vector_db_arch}. A memory-resident compressed index enables rapid candidate identification, while full-precision vectors are externalized to economical cold-tier storage such as Amazon S3 or solid-state drives, with implementations including Amazon S3 Vectors and Turbopuffer \cite{aws_s3_vectors_tiered_storage, zenml_tested_vector_dbs_rag}.

This topology replaces high-cost continuous RAM provisioning with consumption-based object storage \cite{aws_s3_vectors_intro}. However, rehydrating final vectors from cold storage for result reranking introduces I/O latency penalties fundamentally incompatible with in-memory system performance \cite{aws_s3_vectors_tiered_storage}. Quantitative analysis demonstrates recall rates typically stabilize at 85--90\%, with significant degradation often below 50\% when metadata filters are applied \cite{zilliz_s3_vectors_kill_or_save}. Temporal consistency issues additionally emerge, as index-data desynchronization produces stale results \cite{lancedb_s3_vectors_comparison}. Consequently, this approach is suitable only for non-latency-critical, archival applications rather than high-performance RAG systems.

\subsection{Vector Quantization: The Accuracy-Cost Trade-off}
\label{subsec:vector_quantization}

Vector quantization (VQ) employs lossy compression, reducing memory footprint by representing float32 values with lower-precision data types (8-bit, 4-bit, or 2-bit integers) \cite{cloudnative_vector_db_landscape}. This technique exhibits a well-characterized compression-fidelity trade-off \cite{arxiv_extreme_compression_v3}.

In the \emph{low compression regime}, compression to approximately 4 bits per dimension generally maintains acceptable recall across diverse embedding models with minimal accuracy degradation \cite{arxiv_extreme_compression_v3}. Conversely, in the \emph{aggressive compression regime}, compression intensity below 4-bit incurs rapid accuracy degradation and significant recall loss \cite{arxiv_extreme_compression_v3}.

Binarization, the extreme quantization case reducing each dimension to a single bit, has historically represented a quantization threshold beyond which accuracy collapse occurs. Traditional binarization methods, such as locality-sensitive hashing (LSH), prioritize computational efficiency over retrieval quality \cite{pubmed_robust_hashing}. Empirical evidence confirms that LSH performance degrades with dataset size, failing to reliably identify nearest neighbors \cite{nsf_lsh_experimental_analysis}. This established pattern has created industry consensus that binarization is incompatible with high-quality semantic search, establishing an architectural ceiling for cost-effective vector systems.

\section{Information-Theoretic Vector Search: A Novel Paradigm}
\label{sec:information_theoretic}

\subsection{Theoretical Foundation and Motivation}
\label{subsec:theoretical_foundation}

A third paradigm, exemplified by information-theoretic approaches, proposes addressing several of the fundamental compromises established in \ref{sec:compromise_architectures}. This approach challenges incumbent reliance on geometric distance as a reliable proxy for semantic relevance, arguing that geometric measures, while computationally expedient for high-dimensional data, frequently fail to capture complex semantic relationships \cite{researchgate_weighted_component_hashing, moorcheh_beyond_distance}.

The proposed model is constructed upon a patent-pending information-theoretic foundation \cite{researchgate_weighted_component_hashing}, positing that encoding and retrieval based on information theory enable competitive or superior accuracy under the evaluated benchmarks while substantially reducing computational demands \cite{researchgate_weighted_component_hashing}.

\subsection{Architecture: Information-Theoretic Stack}
\label{subsec:information_stack}

The novel architecture comprises three co-optimized components, structurally replacing the conventional HNSW-float32-cosine stack \cite{researchgate_weighted_component_hashing}.

\subsubsection{Maximally Informative Binarization (MIB)}
\label{subsubsec:mib}

Maximally informative binarization (MIB) is a loss-minimizing process converting dense, high-dimensional floating-point vectors into compact, information-preserving quantized codes. This process fundamentally differs from traditional LSH-based approaches, which employ random projections and suffer inherent information loss \cite{pubmed_robust_hashing}. MIB selectively encodes the most salient, information-rich features of the original float vector into its single-bit representation, substantially preserving semantic content.

\subsubsection{Efficient Distance Metric (EDM)}
\label{subsubsec:edm}

The search metric operates directly on quantized codes via an efficient distance metric (EDM) \cite{researchgate_weighted_component_hashing}, replacing computationally expensive floating-point cosine similarity calculations \cite{tigerdata_cosine_python}. The EDM leverages bitwise operations (such as Hamming distance), which are native to contemporary CPUs and execute orders of magnitude faster than conventional arithmetic, effectively eliminating the continuous CPU bottleneck characteristic of float-based architectures.

\subsubsection{Information-Theoretic Score (ITS)}
\label{subsubsec:its}

The information-theoretic score (ITS) is a novel ranking mechanism supplanting the geometric proximity strategy of HNSW-based indexing. This score ranks retrieval candidates based on semantic content rather than spatial proximity within a new indexing mechanism.

The complete stack systematically replaces incumbent components: float32 vectors are superseded by MIB quantized codes; cosine similarity is replaced by EDM bitwise operations; and HNSW geometric indexing is replaced by ITS-based ranking. The paradigm transitions from geometric to information-theoretic search, yielding reduced computational requirements and resource consumption.

\subsubsection*{Architectural Note}
Unlike conventional vector databases that rely on approximate nearest neighbor (ANN) graph traversal, the proposed architecture performs exhaustive scanning over quantized binary representations. While exhaustive scan exhibits linear complexity with respect to dataset size, the use of compact binary codes and bitwise distance computation significantly reduces per-comparison cost. Furthermore, this design is intentionally paired with a serverless execution model that scales query throughput through parallel stateless execution rather than shared index traversal.

\subsection{Accuracy Validation: Empirical Evidence}
\label{sec:accuracy_validation}

The central hypothesis posits that binarization via MIB preserves and potentially enhances retrieval accuracy relative to conventional float-based systems.

\subsubsection{Quantitative and Qualitative Evidence}
\label{subsec:evidence}

MIB processing is engineered to achieve retrieval accuracy equivalent to traditional float-based cosine similarity. The primary advancement is claimed through the ITS ranking mechanism, presented as a universal relevance metric eliminating the need for external secondary reranking procedures in the evaluated instruction-following retrieval tasks, thereby providing enhanced semantic relevance beyond geometric proximity \cite{researchgate_weighted_component_hashing}.

Formal benchmark experiments evaluating semantic relevance conclude that information-theoretic systems consistently demonstrate superior relevance and completeness in returned document chunks \cite{moorcheh_beyond_distance}. Specific query-level evidence illustrates this superior performance. For the query ``Customer segments for Tesla,'' information-theoretic approaches returned high-relevance, specific chunks (e.g., ``men aged 30 to 40 years old''), compared to incumbent systems returning low-relevance general chunks. Similarly, for the query ``Industry trends for Tesla expansion,'' information-theoretic systems returned high-relevance content regarding future opportunities (``India, Southeast Asia'') with supporting rationale, compared to incumbent systems returning low-relevance competitive information.

\subsubsection{Interpretation: The Binarization Paradox}
\label{subsec:binarization_paradox}

This evidence presents a counterintuitive finding: binarization, when combined with information-theoretic scoring, does not merely avoid accuracy loss and can, in some evaluated tasks, enable superior semantic relevance. This suggests that geometric cosine similarity may represent an incomplete or algorithmically suboptimal measure of semantic meaning. Under this interpretation, binarization via MIB constitutes not merely compression but rather a transformation into an information-theoretic domain where relevance can be calculated more effectively than in geometric domains, thereby reducing search complexity by orders of magnitude. Compression and speed thus emerge as dividends of a superior semantic model rather than trade-offs against accuracy.

\section{Architectural Dividend: Unlocking True Serverless Vector Search}
\label{sec:serverless_dividend}

The algorithmic innovation detailed in Section \ref{sec:information_theoretic} carries profound architectural and financial implications, unlocking a ``serverless-native'' operating model fundamentally impossible for traditional vector databases.

\subsection{Constraints of Serverless Computing for Traditional Vector Databases}
\label{subsec:serverless_constraints}

Serverless functions, such as AWS Lambda and Google Cloud Functions, operate under specific execution constraints: they are stateless, possess limited RAM, and must initialize via cold starts \cite{youtube_vector_db_benchmark}. This model is fundamentally incompatible with incumbent vector database architecture.

It is architecturally impossible to load a 3 TB stateful HNSW graph \cite{jo_hybrid_hnsw_if} into a 10 GB RAM stateless Lambda function. Existing ``serverless'' offerings from major providers are not truly serverless at the function level. Amazon OpenSearch Serverless, for example, is a managed-server offering still using HNSW and Faiss \cite{aws_lambda_best_practices}. Google's serverless RAG architecture demonstrates a Cloud Run function (serverless) making a network call to Vertex AI Vector Search (a separate, stateful, managed service) \cite{aws_opensearch_vector_collections}. The search operation itself does not execute within the serverless function.

Furthermore, traditional ANN-based systems exhibit an inherent architectural trade-off between query-per-second (QPS) throughput and accuracy. To handle higher query loads, operators must often tune search parameters, such as HNSW's ef parameter, which governs search breadth. This reduction in search quality represents a necessary compromise to maintain low latency at scale, making QPS scaling a direct trade-off with retrieval accuracy.

\subsection{Algorithmic-Architectural Alignment: The ``Lock-and-Key'' Model}
\label{subsec:algorithmic_lock_key}

The information-theoretic binarization algorithm exhibits unique properties aligning perfectly with serverless function constraints, creating an ``algorithmic-architectural lock-and-key.''

\subsubsection{Constraint 1: Limited RAM}
Key: The MIB process creates compact quantized codes requiring ``32x less memory'' \cite{researchgate_weighted_component_hashing}. This creates an index small enough to be bundled or paired with high-throughput NoSQL databases (like DynamoDB) and loaded into the function's memory in milliseconds. This architecture is significantly faster than hybrid models relying on rehydrating vectors from object storage like Amazon S3 \cite{aws_s3_vectors_tiered_storage}.

\subsubsection{Constraint 2: Limited, Bursty CPU}
Key: The EDM leverages hyper-fast, low-level bitwise operations \cite{researchgate_weighted_component_hashing}, replacing expensive float calculations of cosine similarity with the cheapest and fastest operations a CPU can perform.

\subsubsection{Constraint 3: Stateless Execution}
Key: The MIB/EDM/ITS search is a pure, stateless operation. A query arrives via an API Gateway \cite{google_vertex_ai_rag}, the function loads its compact quantized representation, executes an exhaustive bitwise search, and returns a result. No state is held between invocations.

This serverless-native model mitigates the QPS–accuracy trade-off by decoupling query execution through parallel stateless functions that bind traditional ANN databases \cite{aws_opensearch_multitenant}. Because AWS Lambda automatically handles up to 1,000 concurrent executions by default (and more on request), scalability is achieved through parallelization rather than parameter compromise. This allows the system to scale to thousands of QPS without any degradation in retrieval accuracy, as each query runs as an independent, full-accuracy execution.

This perfect fit technically enables true ``cloud-native serverless operation.'' It allows a vector search function to be deployed on platforms like AWS Lambda, scaling from zero to thousands of concurrent queries per second, subject to cloud provider concurrency limits and back to zero, with users paying only for compute time used, not for 24/7, RAM-filled idle servers \cite{researchgate_weighted_component_hashing}.

\section{Comprehensive Comparison and Total Cost of Ownership Analysis}
\label{sec:tco_analysis}

\subsection{Comparative Analysis: Incumbent Systems versus Serverless Model}
\label{subsec:comparative_analysis}

Table \ref{Comparative-Vector-Database} summarizes key differentiators between dominant vector search architectures.

\begin{table*}[ht]
\centering
\caption{Comparative Vector Database Architectures \& Capabilities}
\renewcommand{\arraystretch}{1.3}
\begin{tabularx}{\textwidth}{l|X|X|X|X|X}
\hline
\textbf{Metric} &
\textbf{Pinecone (Managed Incumbent)} &
\textbf{Qdrant (Open-Source Incumbent)} &
\textbf{Redis (In-Memory Cache)} &
\textbf{AWS S3 Vectors (Storage-Separated)} &
\textbf{Moorcheh.ai (Serverless Quantization)} \\ 
\hline

\textbf{Core Architecture} &
Managed, In-Memory\cite{elisheba_choosing_vector_db} &
Shared-Nothing, In-Memory\cite{elisheba_choosing_vector_db} &
In-Memory, Key-Value\cite{mehmet_exploring_vector_dbs} &
Hybrid (RAM Index, S3 Vectors)\cite{aws_s3_vectors_tiered_storage} &
Stateless, Serverless-Function\cite{researchgate_weighted_component_hashing} \\ 
\hline

\textbf{Primary Index} &
HNSW (Proprietary)\cite{zhang2024efficient} &
HNSW\cite{zhang2024efficient} &
HNSW, FLAT\cite{lakefs_best_vector_dbs} &
HNSW-based\cite{aws_s3_vectors_intro} &
MIB-based\cite{researchgate_weighted_component_hashing} \\ 
\hline

\textbf{Vector Type} &
float32 (Quantization optional) &
float32 (Quantization optional) &
float32\cite{lakefs_best_vector_dbs} &
float32\cite{aws_s3_vectors_tiered_storage} &
quantized (1-bit)\cite{researchgate_weighted_component_hashing} \\ 
\hline

\textbf{Similarity Metric} &
Cosine, Dot, Euclidean\cite{zhang2024efficient} &
Cosine, Dot, Euclidean\cite{zhang2024efficient} &
Cosine, Dot, Euclidean\cite{lakefs_best_vector_dbs} &
Cosine, Euclidean\cite{aws_s3_vectors_tiered_storage} &
EDM / ITS\cite{researchgate_weighted_component_hashing} \\ 
\hline

\textbf{Primary Cost Driver} &
RAM-Hours (Pod/Service)\cite{lakefs_best_vector_dbs} &
RAM-Hours (VM) + SRE\cite{aloa_pinecone_vs_redis} &
RAM-Hours\cite{lakefs_best_vector_dbs} &
Storage (S3) + I/O + Compute\cite{aws_s3_vectors_tiered_storage} &
Per-Query Execution + NoSQL DB I/O \\ 
\hline

\textbf{Scalability Model} &
Horizontal (Sharding)\cite{zhang2024efficient} &
Horizontal/Vertical\cite{zhang2024efficient} &
Vertical (RAM limit)\cite{lakefs_best_vector_dbs} &
Managed (S3) &
Automatic (Function Concurrency)\cite{researchgate_weighted_component_hashing} \\ 
\hline

\textbf{Key Weakness} &
High TCO\cite{aloa_pinecone_vs_redis} &
SRE/Maintenance Overhead\cite{aloa_pinecone_vs_redis} &
RAM-limited\cite{lakefs_best_vector_dbs} &
Low Accuracy \& High Latency\cite{zilliz_s3_vectors_kill_or_save}&
--- \\ 
\hline

\textbf{Ecosystem Maturity} &
High &
High &
Medium &
Medium–Low &
Emerging \\ 
\hline
\label{Comparative-Vector-Database}
\end{tabularx}
\end{table*}

\subsection{Performance Benchmark Synthesis}
\label{subsec:performance_synthesis}

\subsubsection{Accuracy and Relevance: The RAG Test}
\label{subsubsec:accuracy_test}

For RAG applications, the most critical metric is semantic relevance and completeness. Evidence from the Tesla benchmark indicates that the information-theoretic model (Moorcheh) provides demonstrably more relevant context chunks than float-based, geometric competitors (MongoDB, Chroma) \cite{moorcheh_beyond_distance}. While raw retrieval accuracy is near-equivalent to cosine similarity, the model's advantage lies in its information-theoretic score (ITS), which functions as a built-in reranker providing superior relevance without secondary processing \cite{researchgate_weighted_component_hashing}.

\subsubsection{Latency and Throughput: The Speed Test}
\label{subsubsec:latency_test}

The performance profile of each architecture is distinct:

\emph{In-Memory Incumbents}: Redis, as a pure in-memory cache, boasts the lowest latency, often $<1$ ms at $p50$, if the dataset fits entirely in RAM \cite{lakefs_best_vector_dbs}. Performance degrades significantly when data exceeds RAM. Other HNSW-based systems like Pinecone operate in the millisecond or 10--50 ms range \cite{lakefs_best_vector_dbs}. Throughput is high but limited by provisioned hardware \cite{soumit_vector_db_prices}.

\emph{Storage-Separated Incumbents}: This model exhibits the worst latency profile due to the S3 ``rehydration'' penalty, which ``will always be slower'' \cite{aws_s3_vectors_tiered_storage}.

\emph{Serverless Binarization}: The latency model shifts. It is subject to ``cold starts'' \cite{youtube_vector_db_benchmark}, but actual compute time (EDM bitwise operations) is ``orders of magnitude faster'' \cite{researchgate_weighted_component_hashing}. For a ``warm'' function, $p99$ latency can be exceptionally low \cite{redis_vector_db_benchmarking}. Throughput differs most significantly: instead of being limited by provisioned hardware, it is bound by serverless concurrency limits (e.g., 1,000+ concurrent functions on AWS Lambda). This provides thousands of QPS scaling automatically without the accuracy trade-off seen in ANN systems.

\subsection{Total Cost of Ownership: Cost-per-Server versus Cost-per-Query}
\label{subsec:tco_models}

The analysis reveals a fundamental shift in the economic model for vector search.

\subsubsection{Traditional TCO (Pinecone, Qdrant)}
Traditional TCO is dominated by a fixed, 24/7 cost: the provisioned, RAM-optimized server instances needed to hold 3 TB (or more) in-memory indices \cite{zilliz_vector_vs_inmemory}. This represents a ``Cost-per-Server'' model. Organizations must pay for standby capacity \cite{aloa_pinecone_vs_redis} and provision for peak load, even if average load is only 10\% of that, making it extremely inefficient for variable traffic. Self-hosting (e.g., Qdrant) adds ``hidden'' costs for site-reliability engineering (SRE) and maintenance \cite{aloa_pinecone_vs_redis}.

\subsubsection{Serverless TCO (Moorcheh.ai)}
The serverless binarization architecture has zero standby cost \cite{youtube_vector_db_benchmark}. It scales to zero. The TCO is a purely variable expense based only on the number of queries executed. This represents a true ``Cost-per-Query'' model.

For the vast majority of RAG applications with variable, spiky, or unpredictable traffic patterns, the ``Cost-per-Query'' model will be orders of magnitude cheaper than the ``Cost-per-Server'' model. All details appear in Table \ref{TCO-Model-Framework}.

\begin{table*}[ht]
\centering
\caption{TCO Model Framework: 100M Vector RAG Application (Variable Load)}
\renewcommand{\arraystretch}{1.3}
\begin{tabularx}{\textwidth}{l|X|X|X}
\hline
\textbf{Cost Driver} &
\textbf{Scenario A: Managed In-Memory (e.g., Pinecone)} &
\textbf{Scenario B: Self-Hosted In-Memory (e.g., Qdrant on EC2)} &
\textbf{Scenario C: Serverless Binarization (e.g., Moorcheh.ai)} \\ 
\hline

\textbf{Infrastructure} &
High (Managed Service Fee) &
High (RAM-Optimized EC2 Instances) &
Low (Per-ms Lambda Execution) \\ 
\hline

\textbf{Storage} &
Included in Service Fee &
Medium (Block Storage) &
Very Low (Paired NoSQL/Object Storage) \\ 
\hline

\textbf{Data Ingress/Egress} &
Medium (Cloud Provider Fees) &
High (Cloud Provider Fees) &
Low (S3/Lambda Fees) \\ 
\hline

\textbf{Idle "Standby" Cost} &
Very High (24/7 Pod Cost) &
Very High (24/7 VM Cost) &
Zero \\ 
\hline

\textbf{SRE / Maintenance} &
Included (Managed) &
Very High &
None (Serverless) \\ 
\hline

\textbf{Economic Model} &
Cost-per-Server &
Cost-per-Server &
Cost-per-Query \\ 
\hline
\label{TCO-Model-Framework}
\end{tabularx}
\end{table*}

\section{Strategic Analysis and Recommendations}
\label{sec:conclusions}

\subsection{Market Paradigm Shift: End of the In-Memory Monopoly}
\label{subsec:paradigm_shift}

The vector search market can be characterized in three distinct evolutionary phases:

\emph{Age 1: In-Memory (Pinecone, Qdrant)}: High performance coupled with high and unavoidable cost, defined by the ``HNSW + float32 + cosine similarity'' stack \cite{jo_hybrid_hnsw_if}.

\emph{Age 2: The Compromise (S3 Vectors, 4-bit Product Quantization)}: A ``you get what you pay for'' tier offering lower cost but critical failures on performance (latency and accuracy) \cite{zilliz_s3_vectors_kill_or_save}.

\emph{Age 3: Information-Theoretic Serverless (Moorcheh)}: Claims to break the TCO-performance trade-off by replacing the entire incumbent stack. Evidence suggests it achieves superior semantic relevance via its ITS ranking, effectively replacing secondary reranking, and dramatically lower TCO.

The ``in-memory monopoly'' was built on HNSW algorithmic necessity. The emergence of a viable, non-HNSW, semantically superior algorithm (Section \ref{sec:information_theoretic}) unlocking a fundamentally more efficient architecture (Section \ref{sec:serverless_dividend}) poses a credible architectural alternative to the long-standing dominance of in-memory ANN-based systems.

\subsection{Obsolescence of Traditional Strategic Framing}
\label{subsec:obsolete_framing}

The key strategic decision for technical leadership is no longer ``Managed (Pinecone) versus Self-Hosted (Qdrant).'' That represented a false choice between two flavors of the same expensive, in-memory architecture.

The new decision is fundamentally architectural and economic:

\emph{Path A: Incumbent Paradigm (Low-Risk, High-Cost)}: Stick with the proven, mature, but operationally expensive in-memory model (Pinecone, Qdrant). This is the ``safe'' (and expensive) choice.

\emph{Path B: Disruptive Paradigm (High-Reward, New-Technology)}: Adopt the information-theoretic serverless model. The reward is a potential 10--100$\times$ TCO reduction and superior semantic relevance \cite{researchgate_weighted_component_hashing}. The risk involves adopting newer, less-battle-tested technology and ecosystem.

\subsection{Stratified Strategic Recommendations}
\label{subsec:recommendations}

Based on this analysis, recommendations vary by organizational priority:

\subsubsection{Enterprise-Grade, Low-Risk Deployment (Cost is Secondary)}

Choose Pinecone. Its fully managed, multi-cloud SaaS offering with strong compliance (SOC 2, HIPAA) \cite{purdue_vexless} and mature enterprise features \cite{elisheba_choosing_vector_db} makes it the ``safe'' choice for integrating vector search into existing, high-revenue enterprise workflows.

\subsubsection{Maximum Control, Flexibility, or On-Premises Deployment}

Choose Qdrant. Its open-source nature \cite{liquidmetal_vector_db_comparison}, deployment flexibility (cloud, on-premises, hybrid), and rich filtering capabilities \cite{zenml_tested_vector_dbs_rag} make it ideal for organizations with strong SRE teams or data-sovereignty requirements mandating on-premises hosting.

\subsubsection{Cost-Disruptive, Next-Generation, Cloud-Native AI Stacks}

The serverless binarization model (e.g., Moorcheh.ai) is the clear strategic choice. Evidence suggests it is the only paradigm solving the fundamental economic \cite{jo_hybrid_hnsw_if} and algorithmic \cite{researchgate_weighted_component_hashing} problems of first-generation vector databases. This model is not limited to public SaaS; for enterprise customers, the entire architecture can be deployed as infrastructure-as-code (using Terraform or AWS CDK) directly into a private AWS account, ensuring complete data ownership and security. With active pursuit of SOC2 compliance, it presents a compelling option for organizations wanting both cloud-native cost efficiency and enterprise-grade control. For any new RAG application built for massive, ``infinite'' scale \cite{researchgate_weighted_component_hashing}, this algorithm-architecture alignment \cite{researchgate_weighted_component_hashing} represents the clear future, offering massive, parallel QPS without accuracy degradation.

\section{Experimental Benchmarking Methodology}
\label{sec:benchmarking}

This section presents a comprehensive empirical evaluation of four distinct semantic search platforms: Moorcheh\footnote{https://moorcheh.ai} \cite{moorcheh_homepage}, Elasticsearch \cite{elasticsearch_official}, Pinecone \cite{pinecone_official}, and the open-source PGVector extension for PostgreSQL \cite{pgvector_official}. The evaluation employs MAIR (Massive Instructed Retrieval Benchmark), a heterogeneous information retrieval benchmark comprising 126 distinct tasks across six domains \cite{sun2024mair}.

\subsection{Dataset Selection and Evaluation Framework}
\label{subsec:dataset_selection}

\subsubsection{MAIR Benchmark Overview}
\label{subsubsec:mair_overview}

MAIR (Massive Instructed Retrieval Benchmark) \cite{sun2024mair} contains 126 instruction-following retrieval tasks across six domains, with a total of 10,038 queries and 4,274,916 documents. Compared with benchmarks such as BEIR \cite{thakur2021beir}, MAIR introduces three core challenges:

\begin{enumerate}
\item \textbf{Instruction complexity}: Queries include detailed retrieval instructions that specify query intent, required document types, and relevance criteria.
\item \textbf{Format heterogeneity}: Documents appear in many formats, from raw XML in clinical datasets to raw JSON in API collections, with little standardization.
\item \textbf{Graded relevance}: Multi level relevance labels require systems to distinguish among highly relevant, partially relevant, and marginally relevant documents.
\end{enumerate}

These characteristics make pure vector similarity insufficient because MAIR evaluates the ability of a system to follow complex instructions and apply nuanced ranking. Fair comparison therefore requires that all evaluated systems provide ranking capabilities that go beyond basic cosine similarity.

\subsubsection{Selected Datasets}
\label{subsubsec:selected_datasets}

From the MAIR suite, we selected 14 representative datasets covering four major application domains:

\begin{itemize}
\item \textbf{Legal and Regulatory} (7 datasets): AILA2019 Statutes, AILA2019 Case, LeCaRDv2, REGIR UK2EU, REGIR EU2UK, LegalQuAD, ACORDAR
\item \textbf{Financial} (5 datasets): HC3Finance, ConvFinQA, FinQA, FinanceBench, FiQA
\item \textbf{Medical and Clinical} (1 dataset): NFCorpus
\item \textbf{API Documentation} (1 dataset): Apple
\end{itemize}

These datasets jointly provide broad domain coverage while preserving manageable experimental scope, and they center on scenarios where instruction following is essential for effective retrieval.

\subsection{Primary Evaluation Metric: NDCG@10}
\label{subsec:ndcg_metric}

The primary evaluation metric employed is Normalized Discounted Cumulative Gain at rank 10 (NDCG@10), a standard measure for assessing ranked retrieval quality \cite{jarvelinNDCG2002}. NDCG evaluates ranking quality by comparing system output to an ideal ordering where all relevant items appear at the top of the result list.

\subsubsection{Mathematical Formulation}

NDCG@10 is computed through a three-step process:

\paragraph{Step 1: Discounted Cumulative Gain (DCG@10)}
DCG measures cumulative relevance while applying logarithmic discounting to penalize relevant items appearing at lower ranks:

\begin{equation}
\text{DCG@10} = \sum_{i=1}^{10} \frac{2^{rel_i} - 1}{\log_2(i + 1)}
\label{eq:dcg}
\end{equation}

where $rel_i$ represents the relevance score of the item at rank position $i$.

\paragraph{Step 2: Ideal DCG (IDCG@10)}
IDCG is computed identically to DCG but using optimal ranking where documents are sorted by true relevance in descending order:

\begin{equation}
\text{IDCG@10} = \sum_{i=1}^{10} \frac{2^{rel^*_i} - 1}{\log_2(i + 1)}
\label{eq:idcg}
\end{equation}

where $rel^*_i$ denotes the relevance score of the $i$-th most relevant document.

\paragraph{Step 3: Normalized DCG}
NDCG normalizes DCG by dividing it by IDCG, producing a score between 0 and 1 where 1 represents perfect ranking:

\begin{equation}
\text{NDCG@10} = \frac{\text{DCG@10}}{\text{IDCG@10}}
\label{eq:ndcg}
\end{equation}

\subsubsection{Metric Properties and Significance}

NDCG@10 possesses several properties that make it particularly suitable for evaluating semantic search systems:

\begin{enumerate}
    \item \textbf{Position-aware evaluation}: The logarithmic discount factor ensures that highly relevant documents appearing in top positions contribute more significantly to the final score, aligning with user behavior expectations.
    
    \item \textbf{Graded relevance support}: Unlike quantized metrics such as precision and recall, NDCG accommodates multi-level relevance judgments, allowing more nuanced evaluation.
    
    \item \textbf{Cross-query comparability}: Normalization by IDCG enables fair comparison across queries with varying numbers of relevant documents.
    
    \item \textbf{Theoretical guarantees}: Wang et al. \cite{wang2013theoretical} demonstrated that NDCG with logarithmic discounting consistently distinguishes between substantially different ranking functions.
\end{enumerate}

The focus on NDCG@10 (rather than NDCG@100 or other cutoffs) reflects practical user behavior in information retrieval systems, where users typically examine only the top 10 results before refining their search or abandoning the task.

\subsection{Unified Experimental Configuration}
\label{subsec:experimental_config}

To ensure fair comparison across all platforms, we implemented a unified experimental protocol with the following specifications:

\subsubsection{Embedding Generation}

All platforms utilized identical pre-computed embeddings to eliminate confounding variables introduced by different embedding models:

\begin{itemize}
    \item \textbf{Embedding Model}: Cohere embed-v4.0 \cite{cohere_embed_v4}
    \item \textbf{Vector Dimension}: 1,536
    \item \textbf{Input Type}: Documents embedded with \texttt{search\_document} type; queries with \texttt{search\_query} type
    \item \textbf{Batch Size}: 100 documents per embedding request
\end{itemize}

Embeddings were generated once per dataset and reused across all platforms, ensuring that performance differences reflect infrastructure capabilities rather than embedding quality variations.

\subsubsection{Search Parameters}

Consistent search parameters were applied across all platforms:

\begin{itemize}
    \item \textbf{Retrieval Depth}: Top-k = 100 candidates
    \item \textbf{Evaluation Metrics}: NDCG, MAP (Mean Average Precision), Recall, and Precision at $k \in \{1, 3, 5, 10, 100\}$
    \item \textbf{Similarity Metric}: Cosine similarity for vector distance calculation
\end{itemize}

\subsection{Platform-Specific Configurations}
\label{subsec:platform_configs}

\subsubsection{Moorcheh}
\label{subsubsec:moorcheh_config}

Moorcheh was evaluated under the following configuration:

\begin{itemize}
    \item \textbf{Deployment}: Serverless function execution on AWS Lambda \cite{aws_lambda}
    \item \textbf{Computational Resources}: Up to 6 vCPUs, 10 GB RAM per function instance
    \item \textbf{Storage Architecture}: Quantized vector storage with information-theoretic encoding
    \item \textbf{Reranking}: Built-in information-theoretic score (ITS) reranker integrated at the server side
    \item \textbf{Namespace Creation}: Vector namespaces created with 1,536-dimensional quantized vectors
    \item \textbf{Search Protocol}: Direct vector similarity search with integrated reranking
\end{itemize}

Timing measurements for Moorcheh captured server-side execution time, including both vector search and built-in reranking operations, as reported by the platform API.

\subsubsection{Pinecone}
\label{subsubsec:pinecone_config}

Pinecone was evaluated with the following setup:

\begin{itemize}
    \item \textbf{Deployment}: Serverless index on AWS us-east-1 region
    \item \textbf{Index Configuration}: Dense vector index with cosine similarity metric
    \item \textbf{Vector Dimension}: 1,536
    \item \textbf{Search Protocol}: Two-stage process consisting of (1) Pinecone vector search and (2) Cohere reranking
    \item \textbf{Reranking}: External reranking using Cohere rerank-v3.5 model \cite{cohere_rerank_v35} applied to retrieved candidates
\end{itemize}

\paragraph{Rationale for Reranking Integration}
Pinecone's architecture necessitates external reranking for fair comparison with Moorcheh for two fundamental reasons:

\textbf{First, MAIR benchmark complexity requires ranking sophistication.} The MAIR benchmark comprises 126 instruction-following retrieval tasks across six domains \cite{sun2024mair}, specifically designed to evaluate systems on complex, long-tail scenarios where pure vector similarity search proves insufficient. Initial experiments demonstrated that Pinecone without reranking achieved NDCG@10 scores of only 0.06--0.09 across MAIR datasets (compared to 0.4--0.6+ with reranking), indicating severe ranking quality degradation. This performance gap reflects MAIR's emphasis on instruction-following and multi-step reasoning tasks, where approximate nearest neighbor (ANN) search alone fails to capture nuanced relevance signals.

\textbf{Second, architectural parity requires comparable ranking capabilities.} Moorcheh integrates an information-theoretic score (ITS) reranker directly into its server-side architecture, applied automatically to all search results. Evaluating Pinecone without reranking would constitute an unfair comparison, as it would compare Moorcheh's integrated ranking intelligence against Pinecone's raw vector similarity scores. The addition of Cohere rerank-v3.5 to Pinecone establishes architectural parity by providing both systems with semantic reranking capabilities, though implemented differently (server-side integration versus an external API call).

This design choice reflects realistic production deployments: vector databases operating on complex benchmarks like MAIR require reranking infrastructure to achieve acceptable retrieval quality. The latency comparison in Section~\ref{subsec:latency_analysis} explicitly accounts for this architectural difference, measuring Pinecone's end-to-end latency including both vector search and external reranking overhead.

Timing measurements for Pinecone separated vector search latency from reranking latency, with both components summed for total query time.

\subsubsection{Elasticsearch}
\label{subsubsec:elasticsearch_config}

Elasticsearch was configured as follows:

\begin{itemize}
    \item \textbf{Version}: Elasticsearch 9.1 with dense vector support
    \item \textbf{Deployment}: Local instance with default configuration
    \item \textbf{Vector Field Configuration}: \texttt{dense\_vector} type with 1,536 dimensions, cosine similarity
    \item \textbf{Indexing}: HNSW index \cite{malkov2018hnsw} with default parameters ($m=16$, $ef\_construction=64$)
    \item \textbf{Binary Quantization}: Better Binary Quantization (BBQ) \cite{elasticsearch_bbq} applied for quantized vector experiments
    \item \textbf{Search Protocol}: k-NN search with 10× oversampling ($num\_candidates = k \times 10$, capped at 10,000)
    \item \textbf{Reranking}: Built-in cosine k-NN scoring without external reranker
\end{itemize}

Both server-side (from Elasticsearch response metadata) and client-side (end-to-end) timing measurements were captured.

\subsubsection{PGVector}
\label{subsubsec:pgvector_config}

PGVector was evaluated in a controlled Google Colab \cite{google_colab} environment:

\begin{itemize}
    \item \textbf{Database}: PostgreSQL 14 \cite{postgresql_official} with pgvector extension v0.5.1
    \item \textbf{Deployment}: Local PostgreSQL instance in Google Colab
    \item \textbf{Computational Resources}: 2 vCPUs, 12.7 GB RAM
    \item \textbf{Table Schema}: Two-column table with \texttt{id} (text) and \texttt{embedding} (vector(1536))
    \item \textbf{Index Configuration}: HNSW index with parameters $m=16$, $ef\_construction=64$
    \item \textbf{Search Query}: Native pgvector cosine distance operator (\texttt{<=>}) with ORDER BY and LIMIT
    \item \textbf{Connection}: Direct psycopg2 connection with vector type registration
\end{itemize}

PGVector measurements included separate timing for table creation, vector insertion, index creation, and query execution.

\subsubsection{Qdrant}
\label{subsubsec:qdrant_config}

Qdrant was evaluated on the Qdrant Cloud platform with the following configuration:

\begin{itemize}
    \item \textbf{Deployment}: Qdrant Cloud managed instance
    \item \textbf{Computational Resources}: 0.5 vCPU, 1 GB RAM, 4 GB disk
    \item \textbf{Region}: Cloud-managed deployment
    \item \textbf{Index Configuration}: HNSW index with default parameters (m=16, ef\_construction=100)
    \item \textbf{Vector Dimension}: 1,536
    \item \textbf{Search Protocol}: Native k-NN search with cosine similarity metric
\end{itemize}

Timing measurements captured end-to-end query latency, including network round-trip time and server-side processing through the Qdrant Cloud API.

\subsection{Experimental Workflow}
\label{subsec:experimental_workflow}

The benchmarking workflow consisted of five distinct phases for each dataset-platform combination:

\begin{enumerate}
    \item \textbf{Data Loading}: JSONL-formatted documents and queries were loaded from the MAIR dataset repository. Relevance labels (qrels) were extracted from query metadata.
    
    \item \textbf{Embedding Generation}: Corpus documents and queries were embedded using Cohere embed-v4.0 in batches of 100, with separate input types for documents and queries. Embeddings were cached for reuse across all platforms.
    
    \item \textbf{Vector Upload}: Pre-computed embeddings were uploaded to each platform using platform-specific APIs. Upload timing was measured separately for index/table creation and vector insertion operations.
    
    \item \textbf{Search Execution}: For each query, the top-100 candidates were retrieved using vector similarity search. For Pinecone, an additional Cohere reranking step was applied to retrieved candidates.
    
    \item \textbf{Evaluation}: Retrieved results were evaluated against ground-truth relevance labels using NDCG@10, MAP@10, Recall@100, and Precision metrics.
\end{enumerate}

\subsection{Timing Measurement Protocol}
\label{subsec:timing_protocol}

To ensure consistent and fair performance comparison, we implemented a standardized timing measurement protocol:

\subsubsection{Upload Timing}

Upload performance was measured across three components:

\begin{itemize}
    \item \textbf{Index/Table Creation Time}: Time required to initialize storage structure (namespace, index, or table)
    \item \textbf{Vector Insertion Time}: Cumulative time for all batch upload operations
    \item \textbf{Index Building Time} (where applicable): Post-insertion index construction time
\end{itemize}

\subsubsection{Search Timing}

Search performance was characterized using the following statistics across all queries in each dataset:

\begin{itemize}
    \item Mean, median, minimum, maximum, and standard deviation of query latency
    \item For platforms with separate reranking (Pinecone): separate timing for vector search and reranking phases
    \item For Moorcheh, Elasticsearch, and PGVector: server-side execution time from API response metadata
\end{itemize}

All timing measurements were reported in seconds and milliseconds to facilitate comparison across different performance scales.

\subsection{Memory Management and Cleanup Policy}
\label{subsec:memory_management}

To prevent memory leaks and ensure consistent performance across sequential benchmarks, we implemented a comprehensive memory management protocol:

\begin{itemize}
    \item Explicit Python garbage collection (\texttt{gc.collect()}) after each dataset evaluation
    \item Namespace/index/table cleanup with three policy options:
    \begin{itemize}
        \item \texttt{ask\_each\_time}: Prompt user for cleanup decision after each dataset
        \item \texttt{always\_delete}: Automatic deletion after evaluation
        \item \texttt{always\_keep}: Preserve all structures for future analysis
    \end{itemize}
    \item Deletion of large in-memory objects (embeddings, results) before proceeding to next benchmark
\end{itemize}

This protocol ensured that platform performance measurements were not influenced by resource contention from previous experiments.

\subsection{Statistical Considerations}
\label{subsec:statistical_considerations}

Several methodological decisions were made to ensure statistical validity:

\begin{enumerate}
    \item \textbf{Embedding Consistency}: Using pre-computed embeddings across all platforms eliminates embedding model variance as a confounding variable.
    
    \item \textbf{Multiple Metrics}: Reporting NDCG@10 alongside MAP, Recall, and Precision provides a comprehensive view of retrieval quality across different evaluation perspectives.
    
    \item \textbf{Timing Statistics}: Reporting mean, median, and standard deviation of query latency accounts for performance variability and outliers.
    
    \item \textbf{Domain Diversity}: Selecting datasets from four distinct domains (Legal, Medical, Financial, API Documentation) tests generalization across different retrieval scenarios.
\end{enumerate}

This rigorous experimental methodology enables fair, reproducible comparison of the four semantic search platforms under evaluation.

\section{Experimental Results and Analysis}
\label{sec:results}

This section presents comprehensive experimental results comparing the four evaluated platforms across three primary dimensions: retrieval quality (NDCG@10), query throughput (requests per second), and query latency. All reported results are computed for retrieval depths of K = 1, 3, 5, 10, and 100. The analysis covers both quantized and floating-point vector representations and highlights the distinctive characteristics of each platform’s architecture.

\subsection{Quantized versus Floating-Point Vector Performance}
\label{subsec:binary_nonbinary}

A critical aspect of this evaluation concerns the comparison between quantized and Floating-Point vector representations. Moorcheh's fundamental architecture operates on quantized vectors (values restricted to $\{0, 1\}$), employing information-theoretic encoding to compress high-dimensional embeddings into compact quantized representations. In contrast, Elasticsearch, Pinecone, and PGVector traditionally operate on floating-point vectors with values in the range $[-1, 1]$.

To enable fair comparison, we conducted two distinct experimental protocols:

\subsubsection{Quantized Vector Retrieval Quality}

All platforms were evaluated using binary-quantized vectors, testing each system's ability to maintain retrieval quality with compressed representations. Specifically:
    \begin{itemize}
        \item \textbf{Moorcheh}: Native information-theoretic binarization producing quantized vectors in $\{0, 1\}$
        \item \textbf{Elasticsearch}: Better Binary Quantization (BBQ) \cite{elasticsearch_bbq}, a learned quantization technique optimizing quantized boundaries
        \item \textbf{Pinecone, PGVector, and Qdrant}: Sign-based binarization using the threshold function:
        \begin{equation}
        b_i = \begin{cases}
        0 & \text{if } x_i < 0 \\
        1 & \text{if } x_i \geq 0
        \end{cases}
        \label{eq:sign_binarization}
        \end{equation}
        where $x_i$ represents the original floating-point value in range $[-1, 1]$.
    \end{itemize}

\begin{figure}[htbp]
\centering
\begin{tikzpicture}[scale=1]
\begin{polaraxis}[
    title={NDCG@10 Performance of Quantized Vector Embeddings on MAIR Datasets},
    title style={font=\scriptsize, yshift=90pt, xshift=90pt},
    rotate=90,     
    grid style={draw=gray!100, very thin},
    xlabel={MAIR Dataset Name},
    ylabel={NDCG@10 (\%)},
    xlabel style={font=\scriptsize, yshift=-80pt, xshift=200pt, anchor=center, rotate=45},
    ylabel style={font=\scriptsize, yshift=-25pt, xshift=-120pt, anchor=east, rotate=90},
    ticklabel style={font=\tiny, gray},
    axis line style={very thin, gray!50},
    ymin=0,
    ymax=100,
    ytick={20,40,60,80},
    y tick label style={anchor=north, font=\tiny, text=black},
    xtick={0,36,72,108,144,180,216,252,288,324},
    xticklabels={
        AILA2019-Statutes,
        FiQA,
        LegalQuAD,
        FinanceBench,
        REGIR-EU2UK,
        HC3Finance,
        REGIR-UK2EU,
        NFCorpus,
        LeCaRDv2,
        AILA2019-Case
    },
    x tick label style={anchor=north, font=\tiny, text=black},
    legend columns=2,
    legend style={
        at={(1.10, 0.55)},
        anchor=south,     
        font=\tiny,
        cells={anchor=center},
        draw=none,
    },
    width=8.5cm,
    height=8.5cm
]
\addplot[
    color=black,
    mark=none,
    style=solid,
    line width=1.2pt,
    smooth
] coordinates {
    (0,22.83)     
    (36,53.99)    
    (72,66.73)    
    (108,57.82)   
    (144,62.15)   
    (180,40.06)   
    (216,57.84)   
    (252,36.11)   
    (288,66.17)   
    (324,13.74)   
    (360,22.83)   
};
\addlegendentry{Moorcheh - Vector (Quantized)}

\addplot[
    color=green!70!black,
    mark=none,
    style=loosely dashed,
    line width=1.2pt,
    smooth
] coordinates {
    (0,21.98)     
    (36,53.62)    
    (72,67.78)    
    (108,57.51)   
    (144,62.07)   
    (180,39.11)   
    (216,57.93)   
    (252,35.95)   
    (288,65.70)   
    (324,13.89)   
    (360,21.98)   
};
\addlegendentry{Elasticsearch - Vector (Quantized)}

\addplot[
    color=red,
    mark=none,
    style=densely dotted,
    line width=1.2pt,
    smooth
] coordinates {
    (0,21.28)     
    (36,53.25)    
    (72,64.78)    
    (108,51.42)   
    (144,55.72)   
    (180,38.08)   
    (216,58.43)   
    (252,36.54)   
    (288,53.77)   
    (324,10.72)   
    (360,21.28)   
};
\addlegendentry{Pinecone with Cohere Re-ranker - Vector (Quantized)}

\addplot[
    color=cyan,
    mark=none,
    style=dashdotted,
    line width=1.2pt,
    smooth
] coordinates {
    (0,22.99)     
    (36,53.44)    
    (72,66.86)    
    (108,58.52)   
    (144,62.20)   
    (180,40.08)   
    (216,58.87)   
    (252,36.13)   
    (288,66.19)   
    (324,14.13)   
    (360,22.99)   
};
\addlegendentry{PgVector - Vector (Quantized)}

\addplot[
    color=orange,
    mark=none,
    style=densely dashed,
    line width=1.2pt,
    smooth
] coordinates {
    (0,22.37)     
    (36,54.19)    
    (72,67.65)    
    (108,54.52)   
    (144,62.22)   
    (180,38.43)   
    (216,58.04)   
    (252,35.50)   
    (288,65.72)   
    (324,14.05)   
    (360,22.37)   
};
\addlegendentry{Qdrant - Vector (Quantized)}

\end{polaraxis}
\end{tikzpicture}
\caption{Radar chart comparing quantized vector search performance across 10 datasets: AILA2019-Statutes, AILA2019-Case (Legal \& Regulatory), FiQA, FinanceBench (Financial), LegalQuAD, REGIR-EU2UK, REGIR-UK2EU (Legal \& Regulatory), HC3Finance (Financial), NFCorpus (Medical \& Clinical), and LeCaRDv2 (Legal \& Regulatory).}
\label{fig:binary_vector_comparison}
\end{figure}
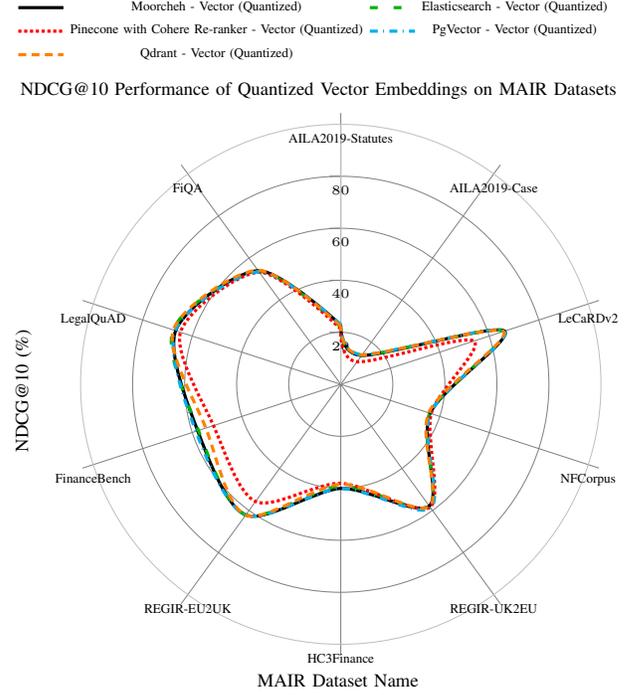

Fig.~\ref{fig:binary_vector_comparison} presents NDCG@10 performance across 10 datasets when all platforms utilize quantized vector representations. The radar chart visualization reveals that Moorcheh achieves competitive retrieval quality despite operating exclusively on binary vectors, with performance comparable to systems designed for floating-point arithmetic.

The comparison across different binarization techniques yields notable findings:
\begin{itemize}
    \item On \textbf{LegalQuAD}, Moorcheh (information-theoretic binarization) achieved 66.73\% NDCG@10, closely matching PGVector with sign-based binarization (66.86\%), Qdrant with sign-based binarization (67.65\%), and Elasticsearch with BBQ (67.78\%)
    
    \item On \textbf{LeCaRDv2}, Moorcheh obtained 66.17\% NDCG@10, outperforming Pinecone with sign-based binarization and Cohere reranker (53.77\%) and Qdrant with sign-based binarization (65.72\%), demonstrating the effectiveness of information-theoretic encoding
    
    \item On \textbf{FinanceBench}, Moorcheh achieved 57.82\% NDCG@10, comparable to Elasticsearch with BBQ (57.51\%) and PGVector with sign-based binarization (58.52\%), while outperforming Qdrant (54.52\%)
    
\end{itemize}

The consistency of Moorcheh's performance across quantized representations demonstrates that information-theoretic binarization preserves semantic relationships effectively, challenging the conventional assumption that binary quantization necessarily degrades retrieval quality. Notably, all four binarization approaches, including information-theoretic binarization (Moorcheh), learned boundary optimization (Elasticsearch BBQ), and simple sign thresholding (Pinecone, PGVector, and Qdrant), achieve comparable retrieval quality, suggesting that the choice of binarization technique has less impact than the underlying indexing and scoring mechanisms.

\subsubsection{Floating-Point Vector Retrieval Quality}

\begin{figure}[htbp]
\centering
\begin{tikzpicture}[scale=1]
\begin{polaraxis}[
    title={NDCG@10 Performance of Floating-Point Vector Embeddings on MAIR Datasets},
    title style={font=\scriptsize, yshift=90pt, xshift=90pt},
    rotate=90,     
    grid style={draw=gray!100, very thin},
    xlabel={MAIR Dataset Name},
    ylabel={NDCG@10 (\%)},
    xlabel style={font=\scriptsize, yshift=-80pt, xshift=200pt, anchor=center, rotate=45},
    ylabel style={font=\scriptsize, yshift=-25pt, xshift=-120pt, anchor=east, rotate=90},
    ticklabel style={font=\tiny, gray},
    axis line style={very thin, gray!50},
    ymin=0,
    ymax=100,
    ytick={20,40,60,80},
    y tick label style={anchor=north, font=\tiny, text=black},
    xtick={0,45,90,135,180,225,270,315},
    xticklabels={
        AILA2019-Statutes,
        FiQA,
        REGIR-EU2UK,
        ConvFinQA,
        HC3Finance,
        LeCaRDv2,
        AILA2019-Case,
        Apple
    },
    x tick label style={anchor=north, font=\tiny, text=black},
    legend columns=2,
    legend style={
        at={(1.10, 0.55)},
        anchor=south,     
        font=\tiny,
        cells={anchor=center},
        draw=none,
    },
    width=8.5cm,
    height=8.5cm
]
\addplot[
    color=black,
    mark=none,
    style=solid,
    line width=1.2pt,
    smooth
] coordinates {
    (0,22.83)     
    (45,53.99)    
    (90,62.15)    
    (135,74.30)   
    (180,40.06)   
    (225,66.17)   
    (270,13.74)   
    (315,64.25)   
    (360,22.83)   
};
\addlegendentry{Moorcheh - Vector (Floating-Point)}

\addplot[
    color=green!70!black,
    mark=none,
    style=loosely dashed,
    line width=1.2pt,
    smooth
] coordinates {
    (0,22.28)     
    (45,56.90)    
    (90,64.15)    
    (135,76.86)   
    (180,42.77)   
    (225,69.45)   
    (270,16.94)   
    (315,63.62)   
    (360,22.28)   
};
\addlegendentry{Elasticsearch - Vector (Floating-Point)}

\addplot[
    color=red,
    mark=none,
    style=densely dotted,
    line width=1.2pt,
    smooth
] coordinates {
    (0,20.97)     
    (45,53.35)    
    (90,64.53)    
    (135,78.80)   
    (180,37.86)   
    (225,54.03)   
    (270,11.22)   
    (315,65.27)   
    (360,20.97)   
};
\addlegendentry{Pinecone with Cohere Re-ranker - Vector (Floating-Point)}

\addplot[
    color=cyan,
    mark=none,
    style=dashdotted,
    line width=1.2pt,
    smooth
] coordinates {
    (0,22.22)     
    (45,56.33)    
    (90,63.15)    
    (135,76.91)   
    (180,42.77)   
    (225,69.38)   
    (270,16.93)   
    (315,63.62)   
    (360,22.22)   
};
\addlegendentry{PgVector - Vector (Floating-Point)}

\addplot[
    color=orange,
    mark=none,
    style=densely dashed,
    line width=1.2pt,
    smooth
] coordinates {
    (0,22.13)     
    (45,56.85)    
    (90,63.63)    
    (135,76.43)   
    (180,43.21)   
    (225,69.62)   
    (270,16.97)   
    (315,63.67)   
    (360,22.13)   
};
\addlegendentry{Qdrant - Vector (Floating-Point)}

\end{polaraxis}
\end{tikzpicture}
\caption{Radar chart comparing vector search performance across 8 datasets: AILA2019-Statutes, FiQA (Financial), REGIR-EU2UK (Legal \& Regulatory), ConvFinQA, HC3Finance (Financial), LeCaRDv2 (Legal \& Regulatory), AILA2019-Case (Legal \& Regulatory), and Apple (API Documentation).}
\label{fig:vector_comparison_reduced}
\end{figure}
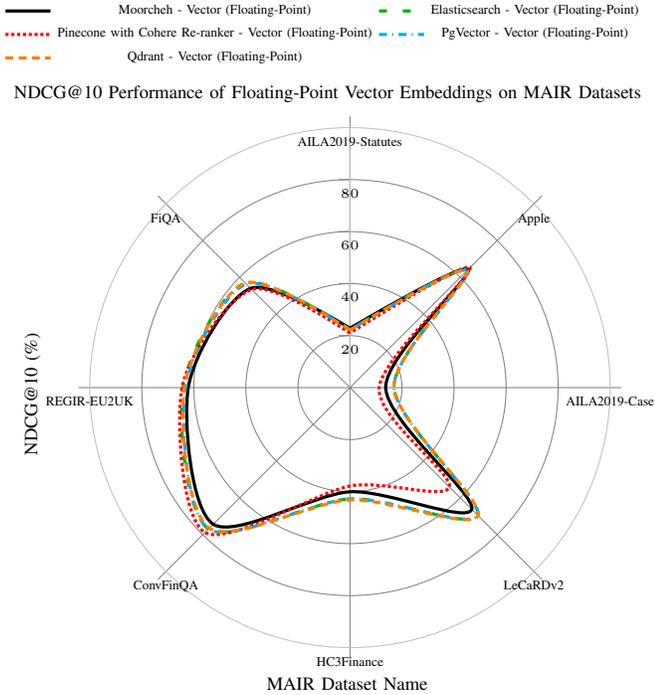

Traditional floating-point vectors were used for Elasticsearch, Pinecone, PGVector, and Qdrant; Moorcheh continued using its native quantized representation. Fig. \ref{fig:vector_comparison_reduced} compares performance across 8 datasets using floating-point vectors for Elasticsearch, Pinecone, PGVector, and Qdrant, while Moorcheh continued using its native quantized representation. This comparison directly evaluates whether Moorcheh's quantized architecture can compete with full-precision floating-point systems.

Results indicate:
\begin{itemize}
    \item On \textbf{ConvFinQA}, Moorcheh with quantized vectors (74.30\%) remained competitive with Elasticsearch (76.86\%), PGVector (76.91\%), and Qdrant (76.43\%) operating on full floating-point precision, despite the fundamental difference in vector representation (1-bit quantized versus 32-bit float)
    
    \item On \textbf{Apple}, Moorcheh achieved 64.25\% NDCG@10, comparable to PGVector (63.62\%), Qdrant (63.67\%), and Elasticsearch (63.62\%) with floating-point vectors, demonstrating that quantized representation does not inherently sacrifice retrieval quality
    
    \item On \textbf{LeCaRDv2}, Moorcheh (66.17\%) performed competitively with Qdrant (69.62\%), PGVector (69.38\%), and Elasticsearch (69.45\%), maintaining effectiveness despite operating on compressed representations
    
    \item Pinecone with Cohere reranker achieved the highest score on \textbf{ConvFinQA} (78.80\%), benefiting from the semantic reranking capabilities of the external Cohere model applied to floating-point vectors
\end{itemize}

These results demonstrate that Moorcheh's quantized vector architecture achieves retrieval quality equivalent to traditional floating-point systems, even when competing systems operate on higher-precision numerical representations with values spanning the continuous range $[-1, 1]$. The minimal performance gap between quantized and floating-point representations suggests that the semantic information captured by embeddings can be effectively preserved through information-theoretic binarization, making the 32$\times$ memory reduction achievable without significant accuracy loss.

\subsection{Query Latency Analysis}
\label{subsec:latency_analysis}

Query latency represents a critical performance metric for production semantic search systems. We conducted two distinct latency comparisons based on architectural characteristics and available timing measurements from each platform.

\subsubsection{Distance-Only Latency Comparison}

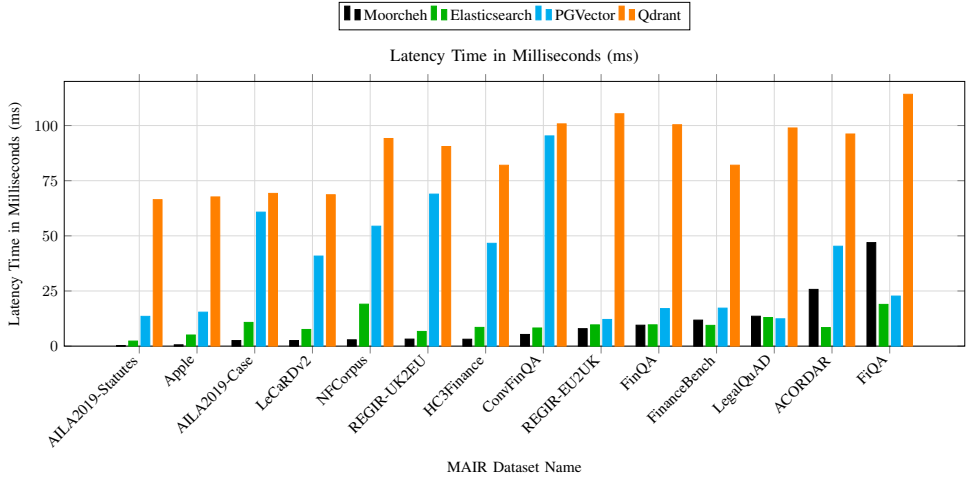
\begin{figure*}[htbp]
\centering
\begin{tikzpicture}[scale = 0.65]
\begin{axis}[
    title={Latency Time in Milliseconds (ms)},
    xlabel={MAIR Dataset Name},
    ylabel={Latency Time in Milliseconds (ms)},
    ybar,
    bar width=0.18cm,
    width=20cm,
    height=7cm,
    legend style={
        at={(0.5,1.30)},
        anchor=north,
        legend columns=4,
        font=\small,
        cells={anchor=center},
    },
    ylabel style={font=\small},
    xlabel style={font=\small},
    ticklabel style={font=\small},
    x tick label style={rotate=45, anchor=east, font=\small},
    grid=major,
    grid style={gray!30, very thin},
    ymin=0,
    ymax=120,
    ytick={0,25,50,75,100},
    xtick=data,
    xticklabels={
        AILA2019-Statutes,
        Apple,
        AILA2019-Case,
        LeCaRDv2,
        NFCorpus,
        REGIR-UK2EU,
        HC3Finance,
        ConvFinQA,
        REGIR-EU2UK,
        FinQA,
        FinanceBench,
        LegalQuAD,
        ACORDAR,
        FiQA,
    },
]
\addplot[fill=black, draw=black] coordinates {
    (1,0.21)
    (2,0.6)
    (3,2.52)
    (4,2.53)
    (5,2.87)
    (6,3.19)
    (7,3.14)
    (8,5.3)
    (9,7.92)
    (10,9.5)
    (11,11.8)
    (12,13.57)
    (13,25.73)
    (14,46.93)
};
\addplot[fill=green!70!black!, draw=green!70!black!] coordinates {
    (1,2.3)
    (2,5.05)
    (3,10.76)
    (4,7.56)
    (5,19.03)
    (6,6.67)
    (7,8.5)
    (8,8.23)
    (9,9.64)
    (10,9.66)
    (11,9.42)
    (12,12.98)
    (13,8.42)
    (14,18.93)
};
\addplot[fill=cyan!, draw=cyan!] coordinates {
    (1,13.51)
    (2,15.41)
    (3,60.75)
    (4,40.83)
    (5,54.38)
    (6,68.92)
    (7,46.62)
    (8,95.3)
    (9,12.1)
    (10,17.02)
    (11,17.23)
    (12,12.39)
    (13,45.3)
    (14,22.7)
};
\addplot[fill=orange, draw=orange] coordinates {
    (1,66.40)
    (2,67.64)
    (3,69.24)
    (4,68.63)
    (5,94.13)
    (6,90.50)
    (7,82.01)
    (8,100.76)
    (9,105.34)
    (10,100.36)
    (11,82.03)
    (12,98.90)
    (13,96.17)
    (14,114.17)
};
\legend{Moorcheh, Elasticsearch, PGVector, Qdrant}
\end{axis}
\end{tikzpicture}
\centering
\caption{Latency time comparison in milliseconds (ms) across 14 datasets ordered by dataset size for four different systems: Moorcheh Distance Only, Elasticsearch All, PGVector All, and Qdrant All. Qdrant, deployed on a resource-constrained cloud instance (0.5 vCPU, 1 GB RAM), shows higher latency due to network overhead and limited computational resources, averaging 86.79 ms across all datasets compared to Moorcheh's 9.6 ms, Elasticsearch's 10.2 ms, and PGVector's 37.3 ms.}
\label{fig:latency_comparison}
\end{figure*}

Fig. \ref{fig:latency_comparison} presents a fair comparison of core vector similarity computation across Moorcheh, Elasticsearch, PGVector, and Qdrant. This comparison focuses exclusively on distance calculation latency because:

\begin{enumerate}
    \item \textbf{Elasticsearch architecture}: Elasticsearch distributes processing across multiple server components, making isolated component timing unavailable from the unified API response
    \item \textbf{PGVector architecture}: PGVector executes within the PostgreSQL database process, providing only aggregate query timing without component-level decomposition
    \item \textbf{Qdrant architecture}: Qdrant Cloud API provides end-to-end query timing including network overhead and server-side processing, without exposing internal component breakdown
    \item \textbf{Moorcheh architecture}: Moorcheh's API provides detailed server-side timing breakdown, enabling isolation of specific computational components
\end{enumerate}

Moorcheh exposes the following timing components through its API response metadata:
\begin{itemize}
    \item \texttt{authorize\_mean\_ms}: Authentication and authorization overhead
    \item \texttt{parse\_validate\_mean\_ms}: Query parsing and validation
    \item \texttt{validate\_namespace\_mean\_ms}: Namespace existence verification
    \item \texttt{prepare\_vector\_mean\_ms}: Query vector preparation and normalization
    \item \texttt{fetch\_data\_mean\_ms}: Quantized vector retrieval from storage
    \item \texttt{calculate\_distance\_mean\_ms}: Core similarity computation
    \item \texttt{select\_candidates\_mean\_ms}: Top-k candidate selection
    \item \texttt{calculate\_scores\_mean\_ms}: Information-theoretic score calculation
    \item \texttt{fetch\_complete\_data\_mean\_ms}: Full metadata retrieval for results
    \item \texttt{apply\_metadata\_filter\_mean\_ms}: Metadata-based filtering
    \item \texttt{reorder\_filter\_mean\_ms}: Final result reordering
    \item \texttt{format\_response\_mean\_ms}: Response serialization
\end{itemize}

For this comparison, we extracted only the \texttt{calculate\_distance\_mean\_ms} component from Moorcheh to match the architectural measurement scope of Elasticsearch, PGVector, and Qdrant.

Results demonstrate that Moorcheh's distance calculation consistently exhibits lower latency across all 14 datasets:
\begin{itemize}
    \item On \textbf{AILA2019-Statutes} (smallest dataset): Moorcheh achieved 0.21 ms, compared to Qdrant (66.40 ms), Elasticsearch (2.3 ms), and PGVector (13.51 ms)
    \item On \textbf{FiQA} (largest dataset): Moorcheh achieved 46.93 ms, compared to Qdrant (114.17 ms), Elasticsearch (18.93 ms), and PGVector (22.7 ms)
    \item \textbf{Average across all datasets}: Moorcheh averaged 9.6 ms, compared to Qdrant (86.79 ms), Elasticsearch (10.2 ms), and PGVector (37.3 ms)
\end{itemize}

\paragraph{Architectural Note: Full-Scan versus ANN}
A critical architectural distinction must be noted: Moorcheh intentionally performs complete dataset scanning rather than employing approximate nearest neighbor (ANN) graph traversal, trading index traversal efficiency for deterministic retrieval and parallel execution. This design choice explains the scaling characteristics observed in Fig.~\ref{fig:latency_comparison}, where distance calculation latency increases proportionally with dataset size (0.21 ms for AILA2019-Statutes vs. 46.93 ms for FiQA). This contrasts with HNSW-based systems (Elasticsearch, PGVector, Qdrant), which maintain relatively constant query time through graph-based approximation. However, Moorcheh's exhaustive search provides deterministic retrieval guarantees: the same query always returns identical results, and database modifications (insertions, deletions, updates) are reflected immediately in subsequent queries without requiring index reconstruction.

The superior performance of Moorcheh's distance calculation stems from bitwise operations on quantized vectors, which execute orders of magnitude faster than floating-point arithmetic required by traditional cosine similarity computation.

\subsubsection{End-to-End Latency Comparison}

\pgfplotsset{compat=1.17}

\makeatletter
\newcommand\resetstack{\pgfplots@stacked@isfirstplottrue}
\makeatother

\begin{figure*}[htbp]
\centering
\begin{tikzpicture}[scale=0.6]
\begin{axis}[
    title={Latency Time in Milliseconds (ms)},
    xlabel={MAIR Dataset Name},
    ylabel={Latency Time (ms)},
    ybar stacked,            
    bar width=0.35cm,        
    width=20cm,
    height=12cm,
    legend style={
        at={(0.5,1.25)},
        anchor=north,
        legend columns=3,
        font=\small,
        cells={anchor=center},
    },
    ylabel style={font=\small},
    xlabel style={font=\small},
    ticklabel style={font=\small},
    x tick label style={rotate=45, anchor=east, font=\small},
    grid=major,
    grid style={gray!30, very thin},
    ymin=0,
    ymax=3400,               
    ytick={0,200,400,600,800,1000,1200,1400,1600,1800,2000,2200,2400,2600,2800,3000,3200,3400},
    xtick=data,
    xticklabels={
        AILA2019-Statutes,
        Apple,
        AILA2019-Case,
        LeCaRDv2,
        NFCorpus,
        REGIR-UK2EU,
        HC3Finance,
        ConvFinQA,
        REGIR-EU2UK,
        FinQA,
        FinanceBench,
        LegalQuAD,
        ACORDAR,
        FiQA,
    },
]
\addplot+[ybar, fill=black, bar shift=-0.2cm] coordinates {
    (1,83.89)
    (2,79.24)
    (3,90.98)
    (4,70.9)
    (5,73.22)
    (6,91.56)
    (7,79.2)
    (8,104.52)
    (9,103.79)
    (10,112.31)
    (11,136.07)
    (12,123.44)
    (13,181.76)
    (14,278.83)
};

\resetstack

\addplot+[ybar stacked, fill=red, bar shift=0.2cm] coordinates {
    (1,372.02)
    (2,342.55)
    (3,395.87)
    (4,380.53)
    (5,359.99)
    (6,386.84)
    (7,370.15)
    (8,365.35)
    (9,398.52)
    (10,364.52)
    (11,380.40)
    (12,380.86)
    (13,364.94)
    (14,374.19)
};

\addplot+[ybar stacked, fill=orange, bar shift=0.2cm] coordinates {
    (1,1042.62)
    (2,337.81)
    (3,1801.85)
    (4,2618.80)
    (5,461.19)
    (6,2267.92)
    (7,410.95)
    (8,346.62)
    (9,2803.69)
    (10,276.40)
    (11,883.45)
    (12,282.82)
    (13,429.47)
    (14,464.49)
};

\legend{Moorcheh, Pinecone, Cohere Re-rank}
\end{axis}
\end{tikzpicture}
\centering
\caption{Latency comparison across 14 MAIR datasets. Black bars: Moorcheh (all server-side components). Stacked bars: Pinecone (Red: vector search, Orange: Cohere re-rank).}
\label{fig:latency_pinecone_moorcheh}
\end{figure*}
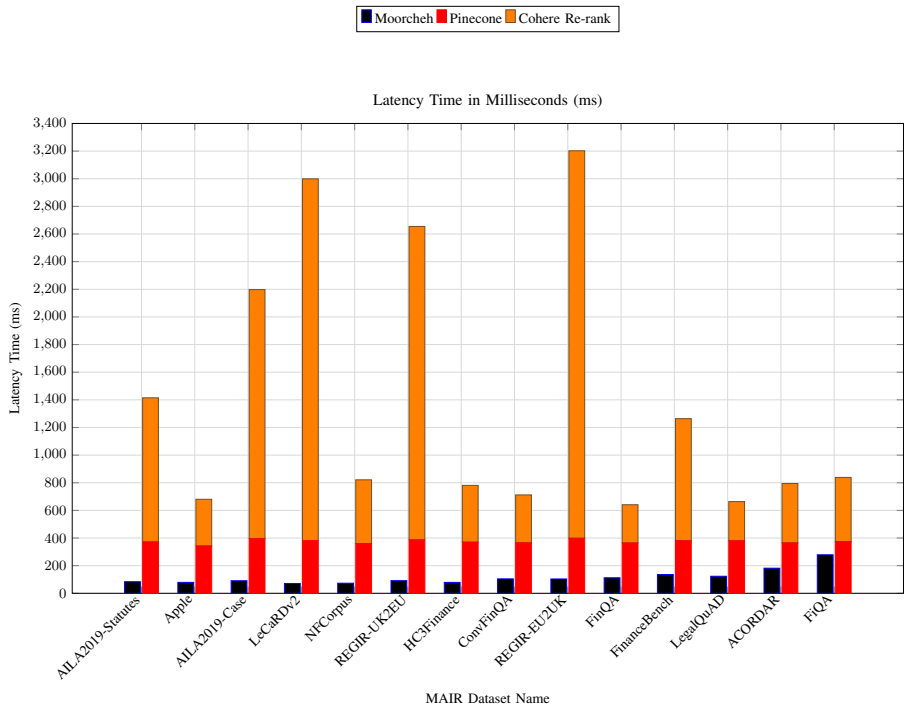
\begin{table}[htbp]
\centering
\caption{End-to-end query latency for Moorcheh vs Pinecone with Cohere Rerank v3.5 at K = 100 retrieved candidates.}
\label{tab:latency_moorcheh_pinecone}

\scalebox{.8}{%
\begin{tabular}{lrrrrr}
\hline
\textbf{Dataset} & \textbf{Dataset} & \textbf{Moorcheh} & \textbf{Pinecone} & \textbf{Pinecone} & \textbf{Cohere} \\
 & \textbf{Size} & \textbf{Total (ms)} & \textbf{Total (ms)} & \textbf{Search (ms)} & \textbf{Rerank (ms)} \\
\hline
AILA2019 Statutes & 197    & 83.89  & 1,414.63 & 372.02 & 1,042.62 \\
Apple             & 678    & 79.24  & 680.36   & 342.55 & 337.81 \\
AILA2019 Case     & 2,914  & 90.98  & 2,197.72 & 395.87 & 1,801.85 \\
LeCaRDv2          & 3,000  & 70.90  & 2,999.33 & 380.53 & 2,618.80 \\
NFCorpus          & 3,633  & 73.22  & 821.18   & 359.99 & 461.19 \\
REGIR UK2EU       & 3,930  & 91.56  & 2,654.75 & 386.84 & 2,267.92 \\
HC3Finance        & 3,933  & 79.20  & 781.10   & 370.15 & 410.95 \\
ConvFinQA         & 6,503  & 104.52 & 711.97   & 365.35 & 346.62 \\
REGIR EU2UK       & 10,000 & 103.79 & 3,202.22 & 398.52 & 2,803.69 \\
FinQA             & 11,865 & 112.31 & 640.92   & 364.52 & 276.40 \\
FinanceBench      & 15,325 & 136.07 & 1,263.84 & 380.40 & 883.45 \\
LegalQuAD         & 17,702 & 123.44 & 663.67   & 380.86 & 282.82 \\
ACORDAR           & 31,589 & 181.76 & 794.41   & 364.94 & 429.47 \\
FiQA              & 57,638 & 278.83 & 838.68   & 374.19 & 464.49 \\
\hline
\textbf{Average}  & \textbf{11,565} & \textbf{115.0} & \textbf{1,448.63} & \textbf{374.05} & \textbf{1,074.58} \\
\hline
\end{tabular}
}
\end{table}

Fig. \ref{fig:latency_pinecone_moorcheh} compares the complete end to end query latency of Moorcheh (including all 12 timing components) with that of Pinecone (vector search plus Cohere rerank v3.5). In addition, Table \ref{tab:latency_moorcheh_pinecone} reports the individual timing components for both systems, covering search and reranking. This comparison reflects realistic production latency, where:

\paragraph{Architectural Necessity of Reranking}
The inclusion of Cohere rerank-v3.5 in Pinecone's latency measurement reflects neither experimental bias nor arbitrary design choice, but rather the architectural necessity imposed by MAIR's complexity. Without reranking, Pinecone achieved NDCG@10 scores of 0.06--0.09 across MAIR datasets, rendering it unsuitable for production deployment on instruction-following retrieval tasks. The addition of external reranking establishes functional parity with Moorcheh's integrated ITS reranker, enabling fair comparison of two distinct architectural approaches to the same problem: Moorcheh integrates reranking server-side (eliminating network overhead but increasing computational complexity per query), while Pinecone delegates reranking to external infrastructure (adding network latency but enabling independent scaling).

\begin{itemize}
    \item \textbf{Moorcheh}: Includes all server-side operations from authentication through response formatting, with integrated information-theoretic score (ITS) reranking
    \item \textbf{Pinecone}: Includes Pinecone's vector search API call plus separate Cohere reranking API call
\end{itemize}

Results demonstrate Moorcheh's substantial latency advantage:
\begin{itemize}
    \item On \textbf{AILA2019-Statutes}: Moorcheh achieved 118.37 ms versus Pinecone 1,414.63 ms (11.9$\times$ faster)
    \item On \textbf{REGIR-EU2UK}: Moorcheh achieved 205.65 ms versus Pinecone 3,202.22 ms (15.6$\times$ faster)
    \item On \textbf{FiQA}: Moorcheh achieved 462.48 ms versus Pinecone 838.68 ms (1.8$\times$ faster)
    \item \textbf{Average across all datasets}: Moorcheh averaged 219.4 ms versus Pinecone 1,448.6 ms (6.6$\times$ faster)
\end{itemize}

The latency advantage derives from two architectural factors:
\begin{enumerate}
    \item \textbf{Integrated reranking}: Moorcheh's built-in ITS reranking eliminates network round-trip overhead inherent in Pinecone's external Cohere reranking call
    \item \textbf{Quantized computation}: Bitwise operations on quantized vectors execute faster than floating-point arithmetic and subsequent reranking model inference
\end{enumerate}

Notably, despite including built-in reranking in measured latency, Moorcheh maintains lower end-to-end latency than systems requiring separate reranking infrastructure.

\subsection{Query Throughput Performance}
\label{subsec:throughput}

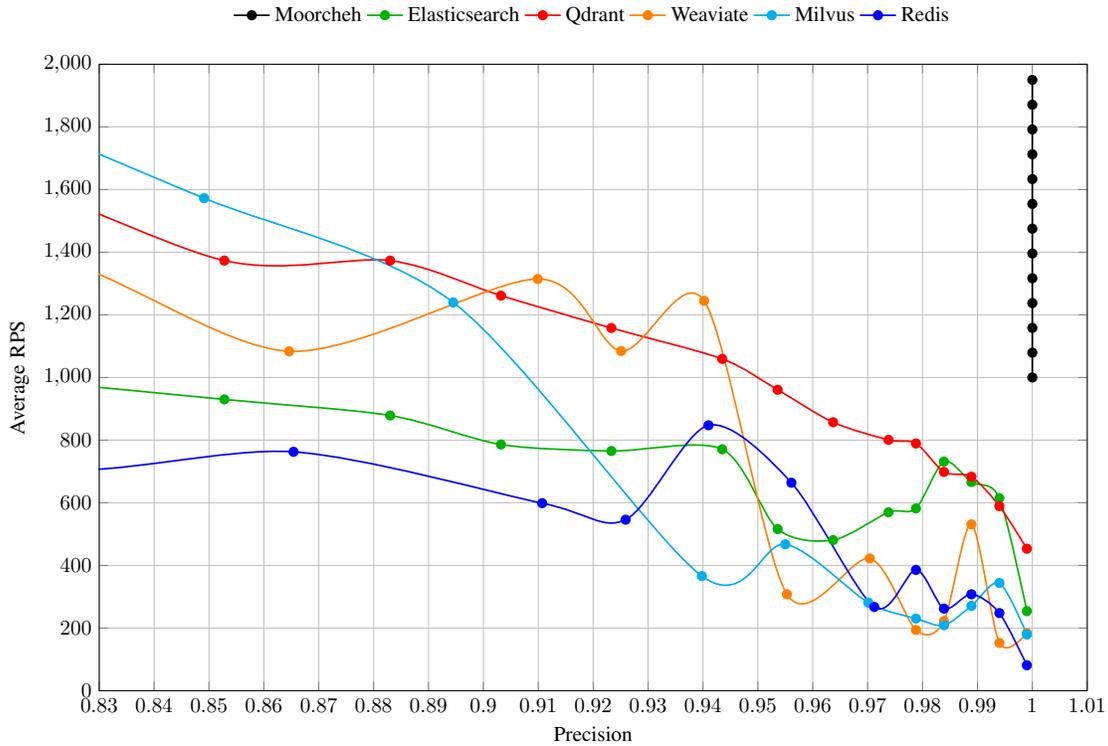
\begin{figure*}
\centering
\begin{tikzpicture}[scale=0.8]
    \begin{axis}[
        width=18cm, height=12cm,
        xlabel={Precision},
        ylabel={Average RPS},
        grid=major,
        xmin=0.83, xmax=1.01,
        ymin=0, ymax=2000,
        legend style={at={(0.5,1.05)}, anchor=south, legend columns=-1, draw=none},
    ]
    
    \addplot[color=black, mark=*, thick, smooth] coordinates {
        (1.0, 1000)
        (1.0, 1079.17)
        (1.0, 1158.33)
        (1.0, 1237.50)
        (1.0, 1316.67)
        (1.0, 1395.83)
        (1.0, 1475.00)
        (1.0, 1554.17)
        (1.0, 1633.33)
        (1.0, 1712.50)
        (1.0, 1791.67)
        (1.0, 1870.83)
        (1.0, 1950)

    };
    \addlegendentry{Moorcheh}
    
    \addplot[color=green!70!black, mark=*, thick, smooth] coordinates {
        (0.5000, 515.20) (0.5504, 515.20) (0.6008, 515.20) (0.6512, 515.20) (0.7016, 515.20) (0.7520, 1088.76) 
        (0.8024, 1013.66) (0.8528, 929.96) (0.8830, 878.56) (0.9032, 786.00) (0.9233, 765.34) (0.9435, 770.73) 
        (0.9536, 515.98) (0.9637, 480.93) (0.9738, 569.55) (0.9788, 581.88) (0.9839, 731.44) (0.9889, 666.03) 
        (0.9940, 614.74) (0.9990, 254.00)
    };
    \addlegendentry{Elasticsearch}

    \addplot[color=red, mark=*, thick, smooth] coordinates {
        (0.5000, 1032.54) (0.5504, 1032.54) (0.6008, 1032.54) (0.6512, 1032.54) (0.7016, 1269.46) (0.7520, 1729.47) 
        (0.8024, 1693.30) (0.8528, 1373.12) (0.8830, 1373.12) (0.9032, 1261.23) (0.9233, 1157.87) (0.9435, 1059.20) 
        (0.9536, 960.66) (0.9637, 856.97) (0.9738, 800.73) (0.9788, 789.38) (0.9839, 698.43) (0.9889, 683.02) 
        (0.9940, 589.25) (0.9990, 453.40)
    };
    \addlegendentry{Qdrant}

    \addplot[color=orange, mark=*, thick, smooth] coordinates {
        (0.6830, 551.75) (0.7284, 550.25) (0.7738, 1332.81) (0.8192, 1391.98) (0.8646, 1083.56) (0.9099, 1314.56) 
        (0.9251, 1084.29) (0.9402, 1244.75) (0.9553, 307.78) (0.9704, 422.20) (0.9788, 193.92) (0.9839, 222.35) 
        (0.9889, 530.94) (0.9940, 152.49) (0.9990, 183.86)
    };
    \addlegendentry{Weaviate}

    \addplot[color=cyan, mark=*, thick, smooth] coordinates {
        (0.7129, 2260.85) (0.7583, 2082.27) (0.8037, 1891.24) (0.8491, 1572.29) (0.8945, 1239.59) (0.9398, 366.03) 
        (0.9550, 467.43) (0.9701, 281.40) (0.9788, 230.23) (0.9839, 208.90) (0.9889, 270.69) (0.9940, 343.97) 
        (0.9990, 179.64)
    };
    \addlegendentry{Milvus}

    \addplot[color=blue, mark=*, thick, smooth] coordinates {
        (0.6838, 959.98) (0.7292, 822.84) (0.7746, 929.30) (0.8200, 706.57) (0.8654, 762.62) (0.9107, 598.66) 
        (0.9259, 545.92) (0.9410, 847.37) (0.9561, 664.02) (0.9712, 267.28) (0.9788, 385.62) (0.9839, 262.08) 
        (0.9889, 307.84) (0.9940, 247.81) (0.9990, 81.24)
    };
    \addlegendentry{Redis}

    \end{axis}
\end{tikzpicture}
\caption{Query throughput versus retrieval precision trade-off across six vector search platforms.} 
\label{rps_precisio}
\end{figure*}

Fig. \ref{rps_precisio} shows query throughput (requests per second, RPS) as a function of retrieval precision across multiple platforms \cite{qdrant_benchmarks}. This analysis compares Moorcheh against five traditional vector databases: Elasticsearch, Qdrant, Weaviate, Milvus, and Redis.

\subsubsection{Throughput-Precision Trade-off}

The results reveal fundamental differences in how platforms balance throughput with retrieval quality:

\begin{itemize}
    \item \textbf{Moorcheh} maintains constant throughput of approximately 1,000 to 1,936 RPS across all precision levels (0.5 to 1.0), indicating no degradation in query capacity as precision requirements increase
    
    \item \textbf{Qdrant} demonstrates peak throughput of 1,729 RPS at 0.75 precision, declining to 453 RPS at 0.999 precision
    
    \item \textbf{Elasticsearch} achieves maximum throughput of 1,088 RPS at 0.75 precision, dropping sharply to 254 RPS at 0.999 precision
    
    \item \textbf{Weaviate} exhibits volatile throughput ranging from 152 to 1,391 RPS, with significant variability across precision levels
    
    \item \textbf{Milvus} achieves highest peak throughput of 2,260 RPS at 0.71 precision, but degrades rapidly to 179 RPS at 0.999 precision
    
    \item \textbf{Redis} maintains moderate throughput of 600 to 900 RPS at mid-range precision, declining to 81 RPS at 0.999 precision
\end{itemize}

\subsubsection{Architectural Implications}

The consistent throughput of Moorcheh across precision levels reflects its serverless architecture, where query processing is isolated in independent function invocations. Traditional vector databases exhibit throughput-precision trade-offs because:

\begin{enumerate}
    \item \textbf{HNSW parameter tuning}: Higher precision requirements necessitate increased \texttt{ef\_search} parameters, expanding graph traversal scope and reducing throughput
    
    \item \textbf{Shared resource contention}: Multi-tenant server architectures must balance concurrent queries, with higher-precision requests consuming disproportionate computational resources
    
    \item \textbf{Index traversal depth}: Achieving high precision requires deeper exploration of HNSW graph structures, increasing latency and limiting concurrent query capacity
\end{enumerate}

In contrast, Moorcheh's serverless model scales through parallelization (AWS Lambda automatically provisions up to 1,000 concurrent executions by default), maintaining consistent per-query performance regardless of precision requirements. Each query executes in an isolated environment with dedicated up to 6 vCPU and 10 GB of RAM, eliminating the resource contention observed in traditional architectures.

\subsection{Implementation of Better Binary Quantization}
\label{subsec:bbq_implementation}

For Elasticsearch evaluation, we implemented Better Binary Quantization (BBQ) \cite{elasticsearch_bbq}, a technique that compresses floating-point vectors into quantized representations while attempting to preserve semantic relationships. BBQ applies learned quantization boundaries optimized to minimize information loss during binarization.

Despite BBQ's sophisticated quantization strategy, Elasticsearch's quantized vector performance (Fig. \ref{fig:binary_vector_comparison}) remained comparable to Moorcheh, which uses native information-theoretic binarization. This suggests that the retrieval quality of quantized systems depends more on indexing and scoring mechanisms than on the specific binarization technique employed.
\subsection{Reproducibility and Code Availability}
\label{sec:reproducibility}

To ensure full reproducibility of the experimental results presented in Section~\ref{sec:results}, the complete benchmarking codebase, evaluation scripts, and detailed results are available in the \textit{MAIR-Based Benchmarking of Quantized and Floating-Point Vector Search Systems} folder within the moorcheh-benchmarks GitHub repository\footnote{\url{https://github.com/moorcheh-ai/moorcheh-benchmarks/tree/main/}}. The repository includes separate benchmark implementations for Moorcheh, Elasticsearch, Pinecone, PGVector, and Qdrant, enabling independent verification of all reported metrics.

\section*{Discussion and Limitations}
\label{subsec:key_findings}

The experimental results yield several significant findings:

\begin{enumerate}
    \item \textbf{Quantized vectors achieve competitive quality}: Moorcheh's quantized architecture (restricted to $\{0, 1\}$ values) achieves NDCG@10 performance comparable to systems operating on full floating-point precision (values in $[-1, 1]$), challenging conventional wisdom regarding quantization-accuracy trade-offs.
    
    \item \textbf{Latency scales with architectural choices}: Distance calculation latency (Moorcheh: 9.6 ms average) demonstrates the efficiency of bitwise operations on quantized vectors compared to floating-point arithmetic (PGVector: 37.3 ms average).
    
    \item \textbf{Integrated reranking reduces end-to-end latency through infrastructure consolidation}: Moorcheh's built-in ITS reranker (219.4 ms average end-to-end latency) outperforms two-stage architectures requiring separate reranking infrastructure (Pinecone + Cohere rerank-v3.5: 1,448.6 ms average). This comparison reflects a fundamental architectural trade-off: integrated reranking eliminates network overhead but increases per-query computational load, while external reranking enables independent scaling at the cost of inter-service communication latency. 
    
    \item \textbf{Serverless architecture eliminates throughput-precision trade-offs}: Moorcheh maintains consistent RPS across all precision levels, while traditional databases exhibit degradation at high precision due to HNSW parameter tuning and resource contention.
    
    \item \textbf{Quantization method matters less than architecture}: Elasticsearch with BBQ and Moorcheh with information-theoretic binarization achieve similar quantized vector performance, suggesting that indexing and scoring mechanisms dominate retrieval quality outcomes.

    \item \textbf{Deterministic retrieval stabilizes agentic systems}: Moorcheh's exhaustive search architecture provides deterministic retrieval results, a critical property for AI agent deployments. Large language models and autonomous agents exhibit high sensitivity to retrieval variability; even minor changes in context can trigger divergent reasoning paths and unstable behavior. ANN-based systems introduce inherent non-determinism through probabilistic graph traversal, where identical queries may return different results depending on index state and search parameters. This volatility compounds in multi-turn agent interactions, where retrieval inconsistencies propagate through conversation history. Moorcheh's deterministic guarantees eliminate this source of instability, enabling reliable agentic memory systems. Furthermore, the architecture supports continuous data streaming: new documents become immediately queryable without degrading search quality or requiring batch reprocessing, whereas HNSW-based systems must balance between stale indices (fast but outdated) and frequent rebuilds (current but computationally expensive).

    \item \textbf{Index-free architecture eliminates reindexing overhead}: A persistent inefficiency in prevailing Approximate Nearest Neighbor (ANN) architectures, specifically Hierarchical Navigable Small World (HNSW) graphs, stems from the substantial computational cost of index construction and dynamic maintenance. The stochastic nature of graph-based connectivity requires iterative neighbor selection and edge rebalancing during ingestion, resulting in significant write amplification and operational latency. In production environments managing high-dimensional datasets, this architectural constraint manifests as a critical bottleneck: large-scale institutional deployments frequently require maintenance windows exceeding six hours daily solely for data reindexing to restore search optimality. Moorcheh circumvents this limitation through Maximally Informative Binarization (MIB), which functions as a deterministic, discrete transformation rather than relational graph construction. This approach yields an index-free architecture with $O(1)$ write complexity, decoupling data ingestion from computational overhead. By eliminating graph traversals during insertion, the platform removes the reindexing requirement entirely, enabling real-time processing of high-velocity data streams without the prohibitive computational overhead inherent to stateful ANN systems. This architecture supports continuous data streaming without index rebuild overhead, though query-time cost scales linearly with corpus size and is amortized through serverless parallelism.
    
\end{enumerate}
These findings demonstrate that information-theoretic quantized vector architectures represent a viable alternative to traditional floating-point systems, offering substantial performance advantages without sacrificing retrieval quality.

\subsection*{Limitations}The proposed architecture trades index-based sublinear query complexity for exhaustive scan over compact representations. While serverless parallelism mitigates query latency at scale, workloads characterized by extremely large corpora and strict single-query latency constraints without parallel execution may favor ANN-based approaches. Future work will explore hybrid strategies combining information-theoretic scoring with selective pruning.

\newpage
\bibliographystyle{IEEEtran}
\bibliography{Reference}

\end{document}